# A Very-High-Velocity Acoustic Wave on a LiNbO$_3$/Sapphire Substrate for Use in Sub-6-GHz Devices

Natalya F. Naumenko, *Member, IEEE*

*Abstract*— This letter presents a numerical investigation of acoustic modes propagating in an LiNbO$_3$/sapphire substrate under a periodic Al grating, demonstrating the occurrence of longitudinal leaky SAWs (LLSAWs) with a unique combination of very high velocities exceeding 10 000 m/s, an electromechanical coupling coefficient of 4.5%, and negligible leakage into the substrate. The leakage was suppressed by optimizing the LiNbO$_3$ (LN) plate and Al electrode thicknesses. Compared with conventional SAW substrates, the LN/sapphire substrate offers 2- to 2.5-fold higher frequencies for periodic electrode structures and enables the production of 6-GHz devices by standard photolithographic processes. Moreover, the required LN plate and Al electrode thicknesses remain compatible with existing wafer bonding technologies. An analysis of LLSAW dispersion in a resonator reveals that the wave behaves as a perfect Rayleigh SAW unperturbed by interactions with spurious modes. The found optimal LN/sapphire substrates are perfect candidates for high-performance SAW devices operating in the sub-6-GHz spectrum.

## I. INTRODUCTION

The deployment of the fifth-generation mobile communication standard (5G), which is characterized by increased network capacity and data rates, has led to an increased demand for high-performance Surface acoustic wave (SAW) filters operating in the sub-6-GHz spectrum. However, the operating frequencies of conventional SAW filters based on LiNbO$_3$ (LN) or LiTaO$_3$ (LT) single-crystal substrates are limited by the minimum line width achievable with state-of-the-art SAW fabrication technologies and thus do not exceed 2.5 GHz. This limitation arises from the velocities of the shear horizontally (SH) polarized acoustic waves (4000–4500 m/s) employed as the primary modes in radio frequency (RF) SAW filters.

This work was partly supported by Ministry of Science and Higher Education of the Russian Federation (Gov. Decree No. 211 of 16 March 2013), National University of Science and Technology "MISIS" (K2-2020-007).
N. F. Naumenko is with National University of Science and Technology "MISIS", Moscow, Russia (e-mail: nnaumenko@ieee.org).

To address this challenge, novel substrate materials that allow the propagation of acoustic waves with much higher velocities are required. Multilayered structures are considered as primary candidates for such substrates. These structures generally consist of a thin piezoelectric plate firmly mounted on a supporting substrate. The plate and substrate are bonded directly or via a few intermediate thin layers to produce an acoustic mirror. Over the last decade, due to the rapid development of breakthrough wafer-bonding technologies, multilayered structures composed of a variety of combined materials have become feasible. The nature and characteristics of acoustic modes propagating in a substrate change with the deposition of additional layers and with variations in the layer thickness. In addition, new higher-order modes appear in such structures with increasing layer thickness. Hence, a variety of SAW device performances can be achieved.

Among the multilayered structures extensively studied as potential SAW substrates [1-5], the most promising substrate is a thin LN or LT plate bonded to an anisotropic substrate. In this case, some of the acoustic modes combine high propagation velocities with the typically large electromechanical coupling coefficients, $k^2$, observed for LN and LT substrates. The high-velocity modes generally demonstrate leaky SAW nature. As a result, SAW resonators employing these modes are characterized by low quality (Q) factors. The leakage can be suppressed by optimizing the anisotropic substrate orientation. For example, a longitudinal leaky SAW (LLSAW) propagating with strong attenuation in an LN substrate can be transformed into a guided plate mode if the LN plate is bonded to a quartz or langasite substrate and if the substrate orientation and plate thickness are optimized. The universal optimization technique combining symmetry consideration of the substrate and plate materials with rigorous simulation of resonator admittances was applied to LT/quartz [6], LN/quartz [7], and LN/langasite [8], and non-attenuated LLSAWs were shown to propagate in these layered structures with velocities of 6000–6500 m/s, rendering these substrates suitable for application in high-frequency high-performance SAW devices. Due to the anisotropy of the substrate, energy confinement of the LLSAW can be achieved without an acoustic mirror between the piezoelectric plate and supporting substrate.

The maximum LLSAW velocity is limited by the longitudinal bulk acoustic wave (BAW) velocity $V_{B3}$ for a specific substrate orientation. LLSAW velocities cannot exceed 7030 m/s and 6830 m/s in LN or LT plates bonded to quartz or langasite, respectively. However, the high velocity limit can be extended up to 11175 m/s if an LN plate is combined with a sapphire substrate.

Although sapphire is less anisotropic than quartz or langasite, the existence of low-attenuated LLSAWs in sapphire with some piezoelectric films has been previously reported. First, non-attenuated LLSAWs propagating with a velocity of 9154 m/s were numerically demonstrated for sapphire with a ZnO film [9]. Recently, low-attenuated LLSAWs with a velocity of 9824 m/s and an $k^2$ up to 1.1% were predicted in sapphire with a ScAlN film, and this phenomenon was confirmed experimentally [10]. Due to the high velocities, these structures are suitable for high-frequency SAW sensors, but their potential application in SAW filters is limited by insufficient $k^2$ values. New bonding technologies enable combining of sapphire with a single-crystal piezoelectric layer. In this letter, the results of a numerical investigation of LLSAWs in LN/sapphire are presented, and the existence of non-attenuated acoustic modes, which combine very high velocities up to 10 500 m/s with $k^2$ reaching 4.5%, is reported.

## II. NUMERICAL RESULTS AND DISCUSSION

The spectrum of acoustic waves propagating in LN/sapphire as the LN thickness is varied in the interval $h_{LN}=(0–0.5\lambda)$ is shown in Fig. 1, where $\lambda=2p$ is the acoustic wavelength and $p$ is the period of the electrode structure. It was obtained numerically for an LN plate defined by the Euler angles (0°, 40°, 0°) bonded to the ZX cut of sapphire. Simulations were performed with periodic Al gratings or electrodes of the interdigital transducer (IDT) located on top of the LN/sapphire structure, as shown in the inset of Fig. 1(a). For the Al electrodes, a finite thickness of $h_{Al}=0.04\lambda$ was applied. For each LN thickness, the resonator admittance function was rigorously simulated using a numerical technique that combines the spectral domain analysis (SDA) of acoustic modes in a layered substrate with a finite-element model (FEM) applied to the electrodes (SDA-FEM-SDA [11]). After the resonant $f_R$ and anti-resonant $f_A$ frequencies were extracted from the simulated admittance functions, the effective velocities of all acoustic modes were estimated at resonance and anti-resonance as $V_R=2pf_R$ and $V_A=2pf_A$, respectively, and the electromechanical coupling coefficients were calculated as $k^2=\pi/2\cdot V_R/V_A\cdot[\tan(\pi/2\cdot V_R/V_A)]^{-1}$. The simulated velocities $V_R$ are shown as functions of LN thickness in Fig. 1(a), and the coupling coefficients $k^2$ of three modes are plotted in Fig. 1(b).

Three zero-order modes can be generated by IDTs when a thin piezoelectric LN plate is added to a non-piezoelectric sapphire substrate. The first mode is a Rayleigh-type SAW, which is most strongly coupled with the electric field. The wave velocity decreases from 5440 to 4037 m/s and $k^2$ increases up to 7% as the LN thickness grows from 0 to 0.5 λ. The second mode arises from a vertically polarized fast shear BAW in sapphire ($V_{B2\text{-sapphire}}$) as a strongly attenuated leaky SAW, but the wave changes its nature and transforms into an SH polarized SAW when its velocity crosses that of the SH-BAW in sapphire ($V_{B1\text{-sapphire}}$). The coupling of this mode remains below 1.4%. The third mode is a strongly attenuated LLSAW arising from $V_{B3\text{-sapphire}}$ at $h_{LN}=0$.

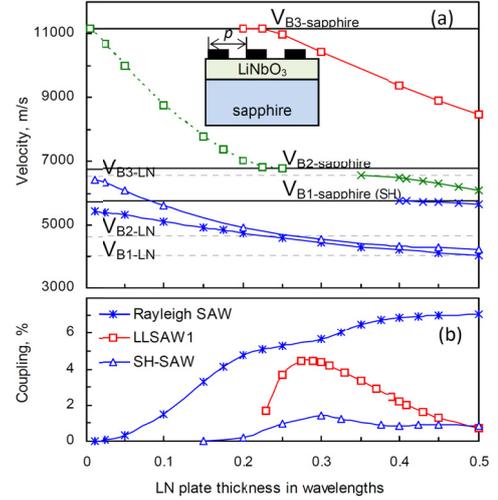

Fig. 1. Velocities (a) and electromechanical coupling coefficients (b) of acoustic modes propagating in (0°, 40°, 0°)-LN/sapphire with $h_{Al}=0.04$ λ, as functions of LN plate thickness. The bulk wave velocities in sapphire and LN are shown for reference.

As the LN plate thickness increases, three higher-order modes arise from $V_{B1\text{-sapphire}}$, $V_{B2\text{-sapphire}}$, and $V_{B3\text{-sapphire}}$, respectively, at $h_{LN}=0.4$ λ, $h_{LN}=0.35$ λ, and $h_{LN}=0.25$ λ. The first-order LLSAW mode (LLSAW1) is in the same velocity interval as the zero-order LLSAW mode, i.e., between $V_{B2\text{-sapphire}}$ and $V_{B3\text{-sapphire}}$, but exhibits much less attenuation caused by leakage into the substrate. Its coupling reaches 4.44% when $h_{LN}=0.28$ λ. In conjunction with velocities varying between 9000 and 11 175 m/s, such $k^2$ renders the LLSAW1 mode highly attractive for use in high-frequency SAW filters. However, the LLSAW leakage requires further suppression via optimization of the plate and electrode thicknesses to enable high resonator Q-factors for RF filter performance.

The attenuation coefficients of leaky waves were estimated from the imaginary components of the complex effective velocities extracted from the admittance functions. To determine the optimal combinations of LN and Al electrode thicknesses ($h_{LN}$ and $h_{Al}$) that offer minimum attenuation of the LLSAW1 mode, the attenuation coefficients were calculated as functions of these two thicknesses. The variations in these coefficients at resonance ($\delta_R$) and anti-resonance ($\delta_A$) as functions of $h_{LN}$ and $h_{Al}$ are shown as contour plots in Figs. 2(a) and 2(b), respectively. The attenuation factor $\delta_R$ vanishes when $h_{LN}=0.27\lambda$ and $h_{Al}=0.04\lambda$, while $\delta_A$ tends to zero when $h_{LN}=0.35\lambda$ and $h_{Al}=0.04\lambda$. Figures 2(c) and 2(d) show the coupling coefficients and velocities of the analyzed waves, revealing that the low-attenuation LLSAWs propagate with $V_R>10\,000$ m/s and $k^2>4\%$.

An example of a simulated admittance function is plotted in Fig. 3(a). Here, the Al electrode thickness was set to its optimal value, $h_{Al}=0.04$ λ, while the LN plate thickness, $h_{LN}=0.3$ λ, was selected to simultaneously achieve low $\delta_R$ and $\delta_A$ values. In the analyzed structure, LLSAW1 propagates with $V_R$=10 444 m/s and $k^2$=4.34%, and the losses transformed into Q-factors yield $Q_R \approx Q_A \approx 950$. The excellent combination of LLSAW characteristics is supplemented by the spurious-free response over a wide frequency range, with only one spurious mode (Rayleigh SAW) observed at $f$=4413 m/s. The nature of the LLSAW1 mode in the optimized structure can be understood from the tangential ($u_1$) and vertical ($u_3$) displacements accompanying LLSAW propagation (colored diagrams shown as insets in Fig. 3 obtained using the numerical technique described in [4]). The LLSAW1 is a sagittally polarized wave, with zero SH displacement, $u_2$=0.

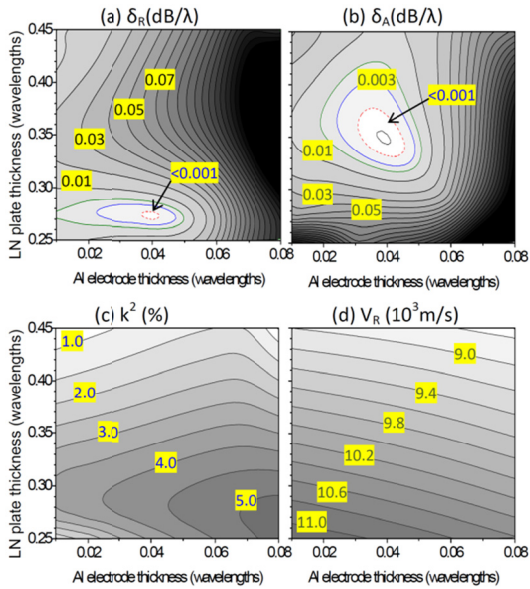

Fig. 2. LLSAW attenuation coefficients at resonance (a) and anti-resonance (b), the electromechanical coupling coefficient $k^2$ (c), and the effective velocity at resonance $V_R$ (d) as functions of Al and LN thicknesses in (0°, 40°, 0°)-LN/sapphire with $h_{Al}$=0.04 λ.

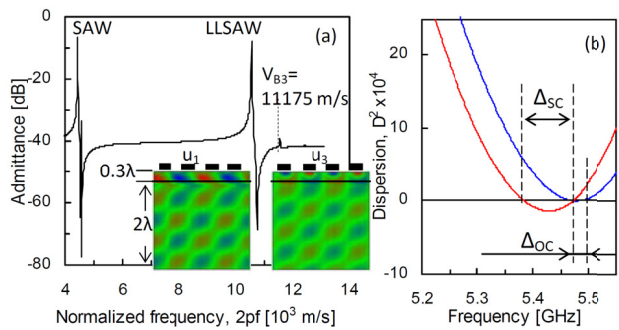

Fig. 3. Simulated admittance of a SAW resonator with Al electrodes on the optimized LN/sapphire structure with contour plots of displacements $u_1$ and $u_3$ illustrating the LLSAW structure (a) and LLSAW dispersion $D^2(f)$ in SC and OC gratings for (0°, 40°, 0°)-LN/sapphire with $h_{LN}$=540 nm, λ=2 μm, and $h_{Al}$=800 Å (b).

To estimate the potential behavior of LLSAWs in a SAW resonator, the squared dispersion $D^2(f)=[(\beta-\beta_0)/\beta_0]^2$, where $\beta_0$ and $\beta$ are the wave numbers unperturbed and perturbed by LLSAW interaction with a periodic grating, respectively, was calculated in short-circuit (SC) and open-circuit (OC) periodic gratings, using the method described in Ref. [12]. For the period of the Al grating $p$=1 μm, the estimated resonator frequencies are $f_R$=5.384 GHz and $f_A$=5.483 GHz. The LLSAW builds a stopband between the frequencies $f_-$=5.384 GHz and $f_+$=5.472 GHz. The parabolic function $D^2(f)$ indicates that the LLSAW does not interact with any spurious modes, and the LLSAW dispersion resembles that of a well-behaved Rayleigh SAW. Hence, the coupling-of-modes (COM) model can be employed to design LLSAW resonators on an LN/sapphire substrate [13]. In the analyzed structure, the LN plate and electrode thicknesses, $h_{LN}$=540 nm and $h_{Al}$=800 Å, remain compatible with the existing wafer-bonding and SAW device fabrication technologies.

In conclusion, the found LN/sapphire substrates enable fabrication of SAW devices with frequencies of up to 6 GHz by standard photolithographic processes and state-of-the-art wafer-bonding technologies.